\documentclass[a4paper]{ar-1col}
\usepackage{comment}
\usepackage{url}
\usepackage[numbers]{natbib}

\jname{Xxxx. Xxx. Xxx. Xxx.}
\jvol{AA}
\jyear{YYYY}
\doi{10.1146/((please add article doi))}

\begin{document}

\markboth{Palmer et al.}{Muon Collider}

\title{Muon Collider: Towards the Energy Frontier}
\author{Mark A. Palmer,$^1$ Daniel Schulte,$^2$ and Vladimir Shiltsev$^3$
\affil{$^1$Brookhaven National Laboratory (retired), Upton, New York 11973, USA}
\affil{$^2$European Organization for Nuclear Research (CERN), CH-1211 Geneva 23, Switzerland; email: daniel.schulte@cern.ch}
\affil{$^3$Northern Illinois University, DeKalb, Illinois 60115, USA}}

\begin{abstract}
A muon collider has emerged as a compelling candidate for the next energy-frontier particle accelerator, combining the discovery potential of multi-TeV lepton collisions with a footprint and power consumption far smaller than existing or proposed hadron and electron-positron colliders. 
Advances over the past two decades in high-power targets, muon production and capture, ionization cooling, rapid acceleration, high-field magnets, collider ring design, and machine-detector integration have transformed the concept from a long-term vision into a credible path toward a future flagship facility. This review summarizes the physics opportunities, the accelerator complex's design and status, the unique experimental environment, recent technical achievements, ongoing international R\&D, and remaining steps toward realizing a high-energy muon collider by the middle of the century.
\end{abstract}

\maketitle

\tableofcontents

\section{INTRODUCTION} 
\label{Introduction}

For more than six decades, particle colliders have been the principal instruments for exploring the fundamental laws of nature~\cite{Shiltsev:2021rmp}. Since the advent of colliding-beam accelerators in the early 1960s, successive generations of electron--positron, electron-proton and proton--(anti)proton colliders have driven virtually every major advance in particle physics, culminating in the establishment of the Standard Model and the discovery of the Higgs boson at the Large Hadron Collider (LHC). Their extraordinary scientific success has established colliders as the most productive facilities for frontier high-energy physics (HEP) research.

That success, however, has come with significantly increasing costs. Progress towards higher collision energies has slowed since the construction of the 14 TeV c.o.m. energy LHC because traditional design approaches have largely exhausted their scaling potential. Circular electron--positron colliders become increasingly constrained by synchrotron radiation, which rapidly grows with energy, while multi-TeV linear colliders require ever greater lengths and complexity. Hadron colliders remain capable of reaching higher energies but demand progressively larger tunnels, stronger superconducting magnets, more elaborate infrastructure, longer construction times, and substantially higher capital and operating costs. Simply building larger, more powerful, and more expensive machines is no longer a sustainable strategy for extending the energy frontier~\cite{roser2023itf}. A fundamentally new paradigm is required and the muon collider (MC) stands out as a uniquely attractive solution~\cite{long2021nature}.

\begin{marginnote}[]
\entry{LHC}{Large Hadron Collider at CERN}
\entry{TeV}{Tera-electron-Volt, 10$^{12}$ eV}
\entry{c.o.m. energy}{Center-of-mass energy, $\sqrt{s}$, the total energy available in the center-of-mass frame of a collision; for two particles of equal energy $E$ colliding head-on, $\sqrt{s}=2E$}
\entry{HEP} {High Energy Physics}
\entry{MC} {Muon Collider}
\end{marginnote}

The muon is an elementary charged lepton, like the electron, but is roughly 200 times heavier. Consequently, synchrotron radiation is suppressed by almost ten orders of magnitude, making it possible to store and collide high-energy muon beams in circular rings. A MC therefore combines two qualities that have previously belonged to different classes of particle colliders: the clean, well-defined collisions of elementary leptons and the energy reach traditionally associated with hadron colliders. It promises multi-TeV collisions in a comparatively compact facility with substantially lower power consumption and potentially lower cost than competing collider concepts. 

The principal challenge is that muons decay into an (anti)electron and two neutrinos $\mu^{\pm} \rightarrow e^{\pm} \nu {\bar{\nu}}$ with a lifetime of only 2.2~$\mu$s at rest. They must be produced as tertiary particles, efficiently captured, rapidly formed into intense low-emittance bunches, accelerated before significant decay occurs, and finally brought into collision. These demanding requirements have stimulated decades of advances in accelerator science and technology, including high-power proton drivers, production targets, pion capture, six-dimensional ionization cooling, rapid acceleration, high-field superconducting magnets, collider ring design, machine--detector interfaces, and mitigation of beam-induced backgrounds arising from muon decay.

The muon collider concept was first proposed by Tikhonin in 1968~\cite{Tikhonin1968MuonBeams} and, independently, by Budker in 1969~\cite{Budker1969Accelerators} and further developed by Skrinsky and Parkhomchuk~\cite{Skrinsky1981Cooling}, Neuffer~\cite{Neuffer1983MuonCooling}, and many others. During the 1990s, studies demonstrated the potential of multi-TeV muon colliders with luminosities exceeding $10^{34}$ cm$^{-2}$s$^{-1}$, resulting in the launch of coordinated international research programs. Over the past three decades, major initiatives — including the first Muon Collider Collaboration~\cite{Ankenbrandt1999MuonRND}, the Neutrino Factory and Muon Collider Collaboration (NFMCC)~\cite{geer2009muon}, the Muon Accelerator Program (MAP)~\cite{Palmer2014MuonColliders, PalmerLong2021MUON}, the Muon Ionization Cooling Experiment (MICE)~\cite{mice2024transverse}, and, currently, the International Muon Collider Collaboration (IMCC)~\cite{accettura2025mcEPPSU} with the closely affiliated US Muon Collider Collaboration (USMCC)~\cite{stratakis2025essay} — have transformed the field. Conceptual designs have matured, key accelerator subsystems have been developed and tested, and technologies once regarded as speculative have advanced to the engineering design phase.
Today, no fundamental technology barriers have been identified that would prevent construction of a multi-TeV MC.

\begin{marginnote}[]
\entry{Luminosity}{a measure of collider performance, $\mathcal{L}$, defined as event rate per reaction cross section; units are cm$^{-2}$s$^{-1}$}
\entry{Integrated luminosity} {$\mathcal{L}_{\rm int}=\int {\cal L}(t) dt$}
\entry{ab} {attobarn, 10$^{-42}$ cm$^2$}
\entry{R\&D} {research and development}
\end{marginnote}
\begin{marginnote}[]
\entry{MAP} {US Muon Accelerator Program, 2010-2017}
\entry{IMCC} {International Muon Collider Collaboration, formed in 2021 and has 65 member institutions}
\entry{USMCC} {The US Muon Collider Collaboration, formed 2025 and is working within the broader scientific framework of the IMCC}
\end{marginnote}

This progress marks an important transition for the field - the central question is no longer whether a MC is physically possible.  Instead the focus is now on how best to optimise its performance, the demonstration of the remaining critical technologies, and integration of these developments into a practical facility. Achieving these goals requires a coordinated international R\&D program, dedicated demonstrator facilities, and continued advances in accelerator, detector, and computation technologies over the next decade.

This review surveys recent progress towards the realization of a MC and key steps that must be accomplished to enable construction of a 3--10~TeV facility around the middle of the twenty-first century. We describe the compelling physics case for a MC, summarize the present status of the accelerator complex and its enabling technologies, discuss the unique experimental environment and detector challenges, and outline the R\&D roadmap towards a Technical Design Report for a future flagship MC facility at the HEP energy frontier.  

High level parameter sets for the target MC energy options, which were developed through the US MAP and IMCC design studies, are presented in \textbf{Table~\ref{tab:accparameters}}.

\begin{table}[!ht]
    \centering

\begin{tabular}{@{}l|c|cc@{}}
\hline
Parameter & Unit of measure & \multicolumn{2}{c}{Target Value} \\
\hline

Center-of-mass Energy $\sqrt{s}$ & TeV & 3 & 10 \\
Interaction Regions & & 2 & 2 \\
Luminosity & $10^{34}$ cm$^{-2}$ s$^{-1}$ & 1.8 & 18 \\
Integrated~Luminosity & ab$^{-1}$ & 1 & 10 \\
\hline
\end{tabular}
 
\caption{Key design parameters for the principal MC
configurations envisioned by the US MAP
and IMCC~\cite{accettura2023towards}.
}
    \label{tab:accparameters}
\end{table}

 
\section{PHYSICS MOTIVATION AND OPPORTUNITIES}
\label{Physics}

A TeV-scale MC combines the energy reach of hadron
colliders with the precision and well-defined initial states
of lepton colliders. Comprehensive reviews of the physics
opportunities offered by MCs can be found in
Refs.~\cite{accettura2025mcEPPSU,smasher2022muon,MuonColliderPhysicsSummary2024}.
Here, we briefly summarize the principal physics motivations
and discovery opportunities that guide the collider design.

\subsection{Energy, luminosity, and the vector-boson fusion regime}

\begin{figure}[hb]
\includegraphics[scale=0.6, width=5in]{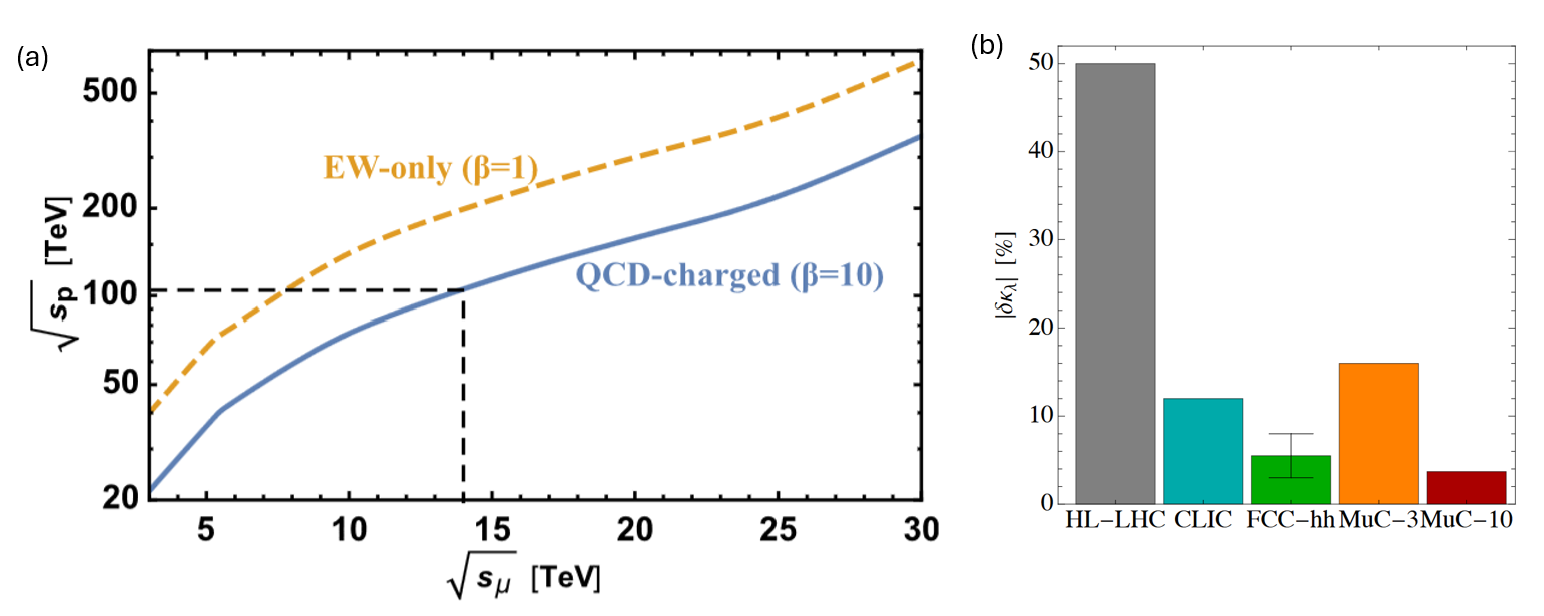}
\caption{a) Energy reach of muon-muon collisions: the energy at which the proton
collider cross section equals that of the MC (from~\cite{wulzer2019muon}).  This compares the pair-production cross-sections for heavy particles with mass $M$ at approximately half the MC energy $\sqrt{s_\mu}$/2. The dashed yellow line assumes comparable processes for muon and proton production while the continuous blue line accounts for the possible QCD enhancement of the production rates at a proton collider, b) Precise measurements of the Higgs boson trilinear coupling will provide deep insight into the shape of the Higgs potential and its connection to the EW phase transition in the early universe. The plot from~\cite{accettura2025mcEPPSU} shows the projected 1$\sigma$ sensitivities, $\delta_{\kappa \lambda}$ (in \%), of various high-energy colliders to this coupling, with the 10 TeV MC providing significant improvement with respect to HL-LHC and the best overall projected precision. Unlike measurements at hadron colliders, the MC measurement precision is expected to be statistically limited.}
\label{fig1}
\end{figure}

Unlike protons, muons are elementary particles, allowing essentially the full c.o.m. energy to participate in hard scattering processes. \textbf{Figure \ref{fig1}a} shows the energies at which cross sections for a proton collider and MC are equivalent for QCD enhanced partonic and Electroweak (EW) processes.  These curves show that a 10-14~TeV MC has an equivalent cross section to a proton collider at $\sim$100~TeV for QCD and $\sim$200~TeV for EW processes. Sensitivities for the Higgs boson trilinear coupling in \textbf{Figure \ref{fig1}b} show the superb relative performance of the MC for the luminosity scenarios in \textbf{Table~\ref{tab:accparameters}}. While these plots do not reflect the full physics potential of a MC, they provide a rough understanding of the relative performance of these two approaches to explore the energy frontier.

At multi-TeV energies, vector-boson fusion (VBF) becomes
increasingly important. While direct $\mu^+\mu^-$ annihilation
cross sections generally decrease as $1/s$, the effective
fluxes of radiated EW bosons grow logarithmically
with energy. Consequently, VBF dominates many production
channels at c.o.m. energies above a few TeV, effectively making the MC 
an EW boson collider
\cite{accettura2025mcEPPSU}.

\subsection{Higgs and Electroweak Physics}


\begin{marginnote}[]
\entry{VBF} {Vector-boson fusion -  a particle physics process in which two incoming quarks or leptons emit virtual electroweak gauge bosons ($W^\pm$ or $Z^0$), which subsequently interact to produce a final state, e.g., a Higgs boson}
\entry{HL-LHC}{High-Luminosity Large Hadron Collider upgrade project, to start operation in 2030}
\entry{CLIC}{ the Compact Linear Collider, a proposed future 3 TeV $e^+e^-$ linear collider concept developed at CERN}
\entry{FCC}{ the Future Circular Collider, a proposed future 91km long facility at CERN, assumes a $e^+e^-$ Higgs factory at the first stage, followed by a 85 TeV proton-proton collider FCC-hh in the same tunnel}
\end{marginnote}


The Higgs sector is a major motivation for a multi-TeV
muon collider. At 10 TeV, an integrated luminosity of
10~ab$^{-1}$ could produce $10^6$--$10^7$ Higgs bosons,
enabling measurements of several Higgs couplings at
sub-percent to percent precision
\cite{Forslund2022,Andreetto2026}.
Double-Higgs production through VBF provides access to
the trilinear Higgs self-coupling $\lambda_3$, with
projected precision improving from about 15--20\%
at 3 TeV to 4--5\% at 10 TeV~\cite{accettura2023towards, Andreetto2026} - see \textbf{Figure \ref{fig1}b}.  This approaches the percent level threshold required to test whether the Electroweak phase transition was first order at the Electroweak scale temperatures.
Triple-Higgs production could also provide sensitivity
to the quartic coupling $\lambda_4$, although the
measurement is considerably more challenging
\cite{Chen2022}.  Additional opportunities include measurements of the
Higgs coupling to muons\cite{Ma2024} and its total decay width\cite{Forslund:2023reu} by exploiting high energy scattering processes or using forward-muon tagging in
$ZZ$ fusion.

High energy vector boson scattering, particularly processes involving longitudinally polarized $W$ and
$Z$ bosons, provides a stringent test of EW symmetry breaking and perturbative unitarity.
Precision measurements of gauge boson, Higgs, and top quark interactions are also sensitive to virtual
effects of heavy new particles. Within the Standard Model Effective Field Theory (SMEFT),
deviations could probe new physics scales of tens of TeV or higher, including Higgs compositeness scales
approaching 100~TeV~\cite{accettura2025mcEPPSU}.  This is a natural consequence of having significant 
luminosity at 10~TeV and low backgrounds, which also extends to flavor physics~\cite{Glioti:2025zpn}.

\subsection{Searches for Physics Beyond the Standard Model}

A multi-TeV MC offers both direct discovery
potential and indirect sensitivity to new physics.
Electroweakly interacting particles can be pair-produced
up to masses approaching half the center-of-mass energy,
while virtual effects and certain single-production
processes may probe substantially higher scales.
Possible targets include additional Higgs bosons,
supersymmetric particles, heavy neutral leptons,
new gauge bosons, and composite resonances
\cite{Liu2024}.

Particularly promising are searches for dark matter candidates belonging to EW multiplets~\cite{Han:2020uak,Capdevilla:2021fmj,Bottaro:2021snn,Han:2021udl}.
Their nearly degenerate charged partners may produce disappearing tracks or other distinctive signatures.
A 10 TeV MC could probe thermal dark-matter benchmarks such as wino-like states near 2.7 TeV and
higgsino-like states near 1.1 TeV, potentially testing much of the parameter space of minimal electroweak
dark matter models. Searches for lepton-flavor violation, extended scalar sectors, and new flavor interactions 
offer further sensitivity to physics beyond the directly accessible mass range.
Additional opportunities for neutrino physics arise from the intense neutrino fluxes produced by muon decays in the collider ring. 

Realizing the projected physics reach of a muon collider will require high luminosity, effective suppression of muon-decay backgrounds, adequate detector acceptance, and precise theoretical predictions. Nevertheless, the combination of direct discovery potential and precision measurements makes a high-energy MC a compelling option for exploring the energy frontier.


\begin{figure}[h]
\includegraphics[scale=0.8, width=5in]{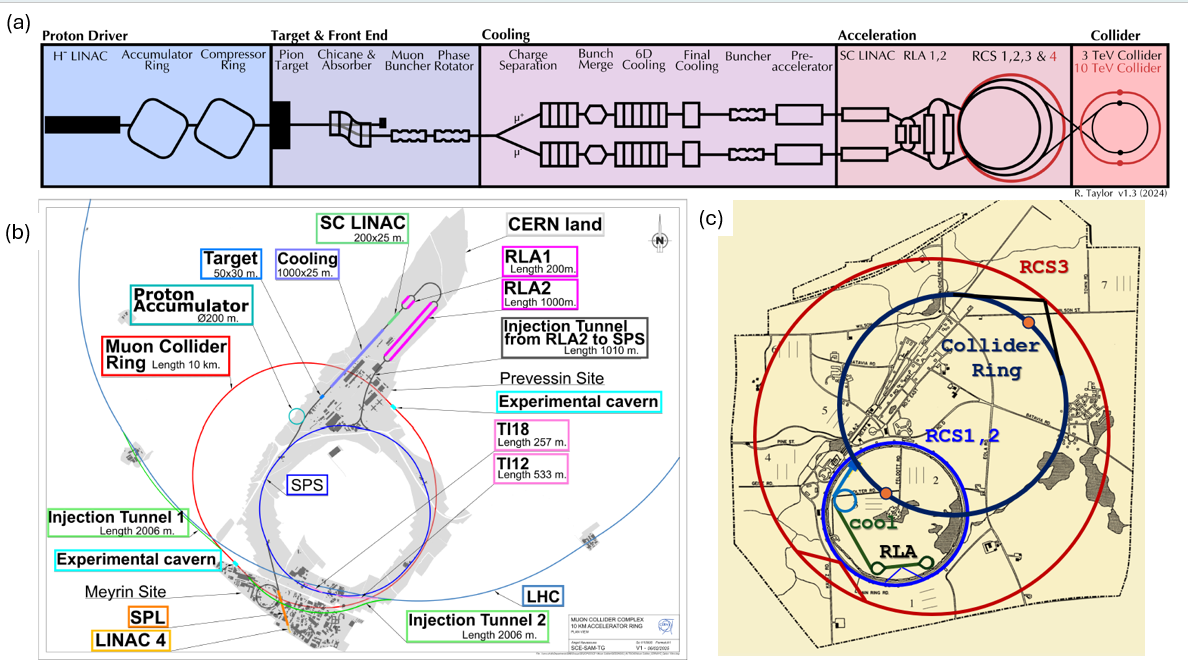}
\caption{Muon collider facility: a) Schematics of the main subsystems of a muon collider, b) Conceptual layout of the muon collider at CERN (left) and Fermilab (right), from ~\cite{accettura2025mcEPPSU}.}
\label{fig2}
\end{figure}

\section{MUON COLLIDER FACILITY AND ACCELERATOR COMPLEX}
\label{Facility}

\subsection{Overall Facility Concept and Parameters}
The concept of the muon collider facility is shown in \textbf{Figure~\ref{fig2}a}. The design of the facility is driven by the short muon lifetime that makes rapid acceleration and a small collider ring essential to obtain high luminosity. The facility operates at $5\;\rm Hz$ and produces, accelerates and collides one $\mu^+$ and one $\mu^-$ bunch per pulse. The main parameters of the 3 and 10 TeV center-of-mass energy MCs are presented in \textbf{Table~\ref{t:facility_param}}.

\begin{table}[!hb]
  \caption{Tentative target parameters for a site independent muon collider with energy staging
    if target performances are achieved. Parameter tables for other designs, some site-specific, at
center-of-mass energies of 0.126, 1.5, 3.2, 6, 7.6, and 14 TeV
can be found in
Refs.~\cite{Palmer2014MuonColliders,neuffer2018feasibility,
boscolo2019future,accettura2025mcEPPSU}, while higher-energy
limitations and luminosity projections are discussed in
Ref.~\cite{Shiltsev2024UltimateColliders}.}
\label{t:facility_param}
\begin{center}

\begin{tabular}{@{}l|c|c|c|c@{}}
    \hline
    Parameter & Symbol & Unit & Stage 1 & Stage 2\\
\hline
    Center-of-mass energy & $\sqrt s$ & TeV & 3 & 10\\
        Beam energy & $E_\mu$ & TeV & 1.5 & 5\\
    Annual integrated luminosity per IP  & $\cal L_{\mathrm{int}}$ & $\rm ab^{-1}$/yr & 0.2 & 2\\
    Luminosity & ${\cal L}$ & $10^{34}\rm cm^{-2}s^{-1}$ & 1.8 & 17.5\\
    Collider circumference& $C_{\mathrm{coll}}$ & $\rm km$ & 4.5 & 11.4\\
    Collider arcs peak field & $B_{\mathrm{arc}}$ & $\rm T$ & 11 & 14\\
    Collider dipole technology& & &$\rm Nb_3Sn$ & HTS\\
    Luminosity lifetime & $N_{\mathrm{turn}}$ &turns& 1039 & 1363\\
    \hline
    Muons/bunch & $N$ & $10^{12}$ & 2.2 & 1.8\\
    Repetition rate & $f_{\mathrm{r}}$ & $\rm Hz$ & 5 & 5\\
    Beam power  & $P_{\mathrm{coll}}$ & $\rm MW$ &5.3  & 14.4\\
    RMS longitudinal emittance& $\varepsilon_\parallel$ & $\rm eVs$ & 0.025 & 0.025\\
    Norm.\,RMS transv. emittance& $\varepsilon_\perp$ & $\mu$m & 25 & 25\\
    \hline
    IP bunch length, rms & $\sigma_z $ & $\rm mm$ & 5 & 1.5\\
    IP beta-function& $\beta $ & $\rm mm$ & 5 & 1.5\\
    IP transverse beam size, rms& $\sigma $ & $\mu$m & 3 & 0.9\\
    \hline
    Protons on target/bunch & $N_{\mathrm{p}}$ & $10^{14}$ & 5 & 5\\
    Protons energy on target  & $E_{\mathrm{p}}$ & $\rm GeV$ & 5 & 5\\
    \hline
  \end{tabular}
\end{center}
\end{table}

The injector produces high-power pulses of $\mathrm{H}^-$ ions. In the accumulator ring, the electrons are stripped off $\mathrm{H}^- + {\rm thin \quad target} \rightarrow p^+ + 2e^-$ , and the resulting protons are accumulated into a single beam pulse. This pulse is then transferred to the compressor ring, where it is compressed to a very short duration. The proton beam strikes a target, producing large numbers of pions that rapidly decay, with charged pions decaying almost exclusively into muons $\pi^{+} \rightarrow \mu^{+} + \nu_\mu$ and $\pi^{-} \rightarrow \mu^{-} + {\bar \nu_\mu}$. The capture system collects the resulting muon beam, which initially has a very large phase-space volume. A sequence of cooling cells reduces its emittance, producing a compact, high-density beam. The muons are then accelerated to their final energy through a sequence of linear accelerators, recirculating linear accelerators, and pulsed synchrotrons. Finally, the collider ring stores the counter-rotating muon beams and brings them into collision twice per turn at two experimental interaction points.

The luminosity of a muon collider can be expressed as~\cite{accettura2023towards}:
\begin{equation}
  \mathcal{L} = f_c\left\langle \frac{N_{\rm coll}^2 }{4\pi \sigma_{\mathrm{IP}}^2} \right\rangle
  = \left( F E_\mu \bar B \sigma_\delta \right)
    \left( \frac{N}{\epsilon_\perp\epsilon_\parallel} \right)
    \left( N E_\mu f_r \right).
\label{lumi}
\end{equation}

The first expression is the standard time-averaged luminosity formula for two identical bunches, each containing $N_{\rm coll}$ particles,  colliding at frequency $f_c$, with an effective transverse rms beam size $\sigma_{\mathrm{IP}}$ at the IP -- see, e.g.,~\cite{Shiltsev:2021rmp}.
For a MC, this expression can be rewritten in terms of a constant factor $F$, the number of injected muons per bunch $N$, the longitudinal and transverse emittances $\epsilon_\parallel$ and $\epsilon_\perp$, the ring's energy acceptance $\sigma_\delta$, and the average magnetic bending field $B$.

\begin{marginnote}[]
\entry{IP}{Interaction Point}
\entry{Emittance}{a beam's phase-space area, characterizing its compactness and quality}
\entry{$\beta$-function} {a focusing beam optics parameter that describes the size of the particle beam $\sigma$ as it travels around the accelerator ring, $\sigma \propto \sqrt{\varepsilon \beta}$}
\end{marginnote}

The factorized form of {\textbf {Equation~\ref{lumi}} highlights several important features of MC performance. The first factor shows that luminosity increases with beam energy $E_\mu$, assuming no additional technological limitations. A higher average bending field $\bar B$ reduces the collider circumference and increases the number of collisions achievable within the muon lifetime, approximately 3000 turns for $\bar B=10$~T. Notably, this number is approximately independent of the beam energy, since both the circumference, $C = 2\pi E_\mu/(0.3\bar B)$, and the relativistically dilated muon lifetime scale linearly with $E_\mu$. A larger relative momentum acceptance $\sigma_\delta$ permits shorter bunches and, consequently, stronger focusing at the IP. The full expression for the constant $F$ is given in~\cite{accettura2023towards}.

The second factor characterizes the muon beam phase-space density through the bunch population $N$ and the transverse and longitudinal emittances, $\epsilon_\perp$ and $\epsilon_\parallel$, respectively. The third factor, $NE_\mu f_r$, represents the average beam power injected into the collider ring, where $f_r$ is the repetition rate.

An important consequence is that the luminosity per unit beam power increases linearly with beam energy, provided the other factors remain approximately constant. For the nearly constant bunch population and repetition rate assumed in \textbf{Table~\ref{t:facility_param}}, the luminosity scales approximately quadratically with collision energy, whereas beam power increases only linearly. This favorable scaling of luminosity per unit beam power is a distinctive advantage of muon colliders compared with other high-energy collider concepts.

\subsection{Proton Driver}

The injector linac accelerates $\mathrm{H}^-$ ions to $5\mathrm{GeV}$ at a repetition rate of 5 Hz, delivering $5\times10^{14}$ ions per pulse and a total beam power of approximately $2\mathrm{MW}$. Each pulse consists of a train of bunches that are accumulated in the accumulator ring using charge-exchange injection. The negative charge of the incoming ions allows them to be merged with the circulating proton beam, after which their two electrons are stripped off.

Once the full pulse has been accumulated, the proton beam is transferred to the compressor ring, where an RF system introduces an energy chirp that enables compression to a pulse duration of approximately 2 ns. The compressed pulse is then directed onto the pion production target.


\begin{marginnote}[]
\entry{SNS} {Spallation Neutron Source at ORNL, USA, a 2 MW 1.3 GeV pulsed SRF proton linac, in operation since 1999}
\entry{ESS}{European Spallation Source in Lund, Sweden, a multi-MW 2.0 GeV pulsed SRF proton linac, beam operation started 2025 }
\entry{SPL} {Super Proton Linac, a CERN design study of a 4 MW 4 GeV pulsed SRF proton linac}
\entry{RF}{radio-frequency cavities and systems that use high-frequency electromagnetic fields, typically from hundreds of MHz to a few GHz, to transfer energy to charged particles}
\entry{SRF}{Superconducting RF cavities operating at cryogenic temperatures to minimise RF power losses}
\entry{HTS}{High-Temperature Superconductors, a new class of conductors for accelerator magnets that can operate at temperatures significantly above the $\sim$2--4 K required for conventional superconductors (NbTi or Nb$_3$Sn)}
\entry{FLUKA}{a general-purpose, fully integrated Monte Carlo simulation package to calculate particle transport and radiation interactions with matter \cite{FLUKA}}
\entry{chicane} {a subsystem of several dipole magnets—typically four—used to alter the path of a particle beam}
\end{marginnote}

Conceptual designs have been developed for the major components of the proton driver complex. The injector linac follows established hadron linac designs, such as those of SNS and ESS and the SPL study, and relies on well-proven technologies. Detailed simulation studies have established the designs of the more challenging accumulator and compressor rings and demonstrated their ability to meet the required performance specifications.

\subsection{Muon Production and Target Systems}
The 5 GeV proton beam strikes a graphite rod target, producing large numbers of pions that rapidly decay into muons. Each proton pulse deposits a substantial fraction of its $400\,\mathrm{kJ}$ of energy in the target, generating intense thermal and mechanical shocks.

A graphite target has been designed and optimised for these demanding conditions. Detailed simulations indicate that it can withstand the resulting thermal shocks and sustained irradiation at the full beam power of $2\,\mathrm{MW}$. Alternative target concepts are also being investigated to accommodate potentially higher beam powers and pulse energies. In particular, liquid-metal jets and tungsten-powder jets are promising candidates.

The target is surrounded by a high-field solenoid that captures the secondary particles and guides them along helical trajectories into the downstream channel. A high-temperature superconducting (HTS) solenoid has been designed to provide the required peak magnetic field of $20\,\mathrm{T}$ at the target, enabling efficient capture of pions and muons. Its large aperture accommodates substantial shielding against secondary particles and radiation produced by the proton beam striking the target. Detailed FLUKA simulations have been used to optimise this shielding, reducing heat deposition and radiation exposure in the superconducting coils to acceptable levels.

A tapered-field solenoid system transports the captured particles through a decay channel, where most of the remaining pions decay into muons. A subsequent chicane separates the muons from residual protons and other unwanted particles, which are intercepted by absorbers. An RF system then captures the muons longitudinally and organizes them into a short train of bunches. Finally, the positive and negative muons are separated into two beams, each directed to its own cooling system.

Early MC design studies explored the possibility of using polarized beams~\cite{Ankenbrandt1999MuonRND}. Although captured muons have a natural average polarization of approximately 20\%, higher polarization can be achieved through selection based on pion decay kinematics, albeit at the cost of a substantial reduction in muon intensity. Moreover, preserving polarization throughout the accelerator chain introduces significant additional complexity. Consequently, current MC designs do not assume polarized beams at collision.

\begin{figure}[h]
\includegraphics[width=5in]{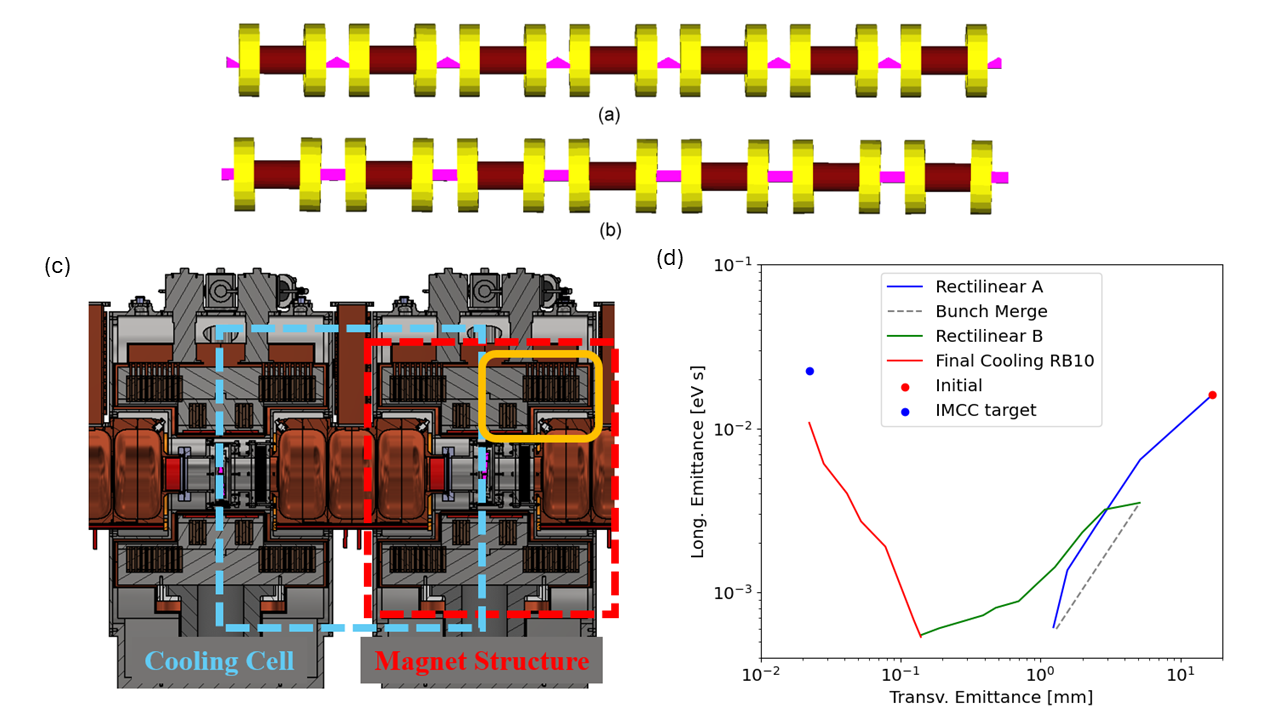}
\caption{Muon beam cooling: a) Conceptual design of a rectilinear muon ionization cooling channel: top view and side view. The large yellow cylinders are solenoids, the small red cylinders are the active volume of the RF cavities, and the magenta wedges are hydrogen absorbers;  b)  side view of one cell of the last stage (stage A4), from~\cite{stratakis2015rectilinear}; c) The overall performance estimate of the MC ionization cooling system: the beam initial emittance on the right and the reduction in both emittances
via the 6D rectilinear cooling (green), further reduction in transverse emittance comes at the cost of longitudinal emittance via final cooling (orange)~\cite{accettura2025mcEPPSU}.}
\label{fig3}
\end{figure} 

\subsection{Ionization Cooling}

Since muons are produced as tertiary particles through the process $p + \mathrm{target} \rightarrow \pi^{\pm} \rightarrow \mu^{\pm}$, the resulting beams exhibit large energy and angular spreads, corresponding to large longitudinal and transverse emittances, $\epsilon_\parallel$ and $\epsilon_\perp$, respectively. These large emittances limit the efficiency of beam capture and acceleration in the downstream accelerator chain and prevent the collider from achieving the required high luminosity (see \textbf {Equation \ref{lumi}}). Beam cooling, which reduces the beam emittances and transverse dimensions, is therefore essential.

Moreover, the short muon lifetime requires exceptionally rapid cooling. Conventional cooling techniques are too slow to be effective, leaving ionization cooling as the only practical approach for muon beams. The muon cooling system consists of a sequence of absorbers and RF accelerating cavities embedded in a magnetic focusing lattice, as illustrated in \textbf{Figures \ref{fig3}a and \ref{fig3}b}. 
As muons traverse the absorbers, they lose both transverse and longitudinal momentum through ionization energy loss, while the RF cavities subsequently restore only the longitudinal component. Consequently, the transverse momentum and particle trajectory angles are reduced. Repeated application of this process decreases the beam angular spread and transverse emittance, producing a more compact beam.

Strong solenoids around the absorbers and cavities guide the muons and reduce the impact of the muons scattering on the nuclei of the absorbers which otherwise would dilute the beam again. The evolution of the normalized transverse emittance is approximately described by~\cite{Skrinsky1981Cooling, Neuffer1983MuonCooling}:
\begin{equation}
  d\epsilon_\perp/ds \approx -\frac{\epsilon_\perp}{\beta_\mu^2 E_\mu}\left|\frac{dE}{ds}\right|
  + \frac{\beta_\perp (13.6\,\mathrm{MeV})^2}{2\beta^3 E_\mu m_\mu c^2 L_R}
  \label{cooling}
\end{equation}
Here, $\beta_\perp$ is the transverse lattice beta function, which, for continuous solenoidal focusing with an axial magnetic field $B_z$, is approximately $\beta_\perp \simeq 2\beta_\mu \gamma_\mu m_\mu c/(eB_z)$. The relativistic factors are $\beta_\mu=v_\mu/c$ and $\gamma_\mu=E_\mu/(m_\mu c^2)$. The first term describes cooling due to ionization energy loss, $dE/ds$; the second represents emittance growth (heating) caused by multiple small-angle Coulomb scattering from the absorbers nuclei and contains the radiation length of the media $L_R$.

Balancing the cooling and heating terms yields the equilibrium transverse emittance. For solenoidal focusing, its approximate dependence is
\begin{equation}
\epsilon_{\rm eq}(E_\mu)\propto \frac{1}{L_R\left.\frac{dE}{ds}\right|_{E_\mu}}\frac{1}{B_z}\gamma_\mu
\label{equilibriumemittance}
\end{equation}
The equilibrium emittance is therefore minimised by using absorber materials with large ionization energy loss per radiation length and strong focusing fields. Hydrogen is particularly advantageous because of its favorable combination of high stopping power and long radiation length, while higher solenoidal fields further reduce the minimum achievable emittance. 

Transverse ionization cooling can operate over a wide range of muon momenta, but its effectiveness depends strongly on energy. At low momenta, the negative slope of the ionization energy-loss curve, $dE/ds$, causes lower-energy muons to lose more energy than higher-energy ones, increasing the longitudinal emittance. At high momenta, the RF power required to replenish the lost energy becomes excessive. Consequently, the initial cooling stages typically operate at momenta of approximately $200$--$300\,\mathrm{MeV}/c$, near the minimum-ionizing region.

The Muon Ionization Cooling Experiment (MICE) at Rutherford Appleton Laboratory (RAL), shown in \textbf{Figure~\ref{fig4}c}, provided the first experimental demonstration of muon ionization cooling. By passing a $140\,\mathrm{MeV}/c$ muon beam through low-$Z$ absorbers, namely liquid hydrogen ($\mathrm{LH}_2$) and lithium hydride ($\mathrm{LiH}$), under strong focusing by superconducting solenoids, MICE demonstrated a reduction in transverse emittance and a corresponding increase in phase-space density, as illustrated in \textbf{Figure~\ref{fig4}d}. The experiment demonstrated cooling in a single absorber stage, without RF reacceleration, which will be essential for sustained cooling in a multi-stage system.

A complete cooling system has been designed and its performance evaluated through simulations~\cite{boscolo2019future,accettura2025mcEPPSU}. The evolution of the transverse and longitudinal emittances through the successive cooling stages is illustrated in \textbf{Figure~\ref{fig3}c}. The incoming muon beams initially have large emittances in both planes (upper right corner). The first stages provide six-dimensional (6D) cooling, combining transverse ionization cooling with longitudinal emittance reduction. Operating near the minimum of $dE/ds$ limits the growth of the energy spread, while weakly bending the beam introduces dispersion, allowing wedge-shaped absorbers to reduce the longitudinal emittance through emittance exchange.
Following the initial cooling stage, the muon bunch train is merged into a single bunch, which undergoes further 6D cooling.

The underlying physics of muon energy loss and longitudinal emittance reduction using wedge absorbers has been experimentally demonstrated in the Muon $g-2$ experiment~\cite{stratakis2019wedge,muongminus22024detailed}. Recent advances in cooling system design have achieved substantial reductions in both transverse and longitudinal beam emittances in simulations, meeting the collider design targets. 
Advanced artificial intelligence and machine-learning (AI/ML) methods offer further opportunities to improve muon cooling channel performance~\cite{AIML2026ai4muc}.

\begin{figure}[h]
\includegraphics[width=5in]{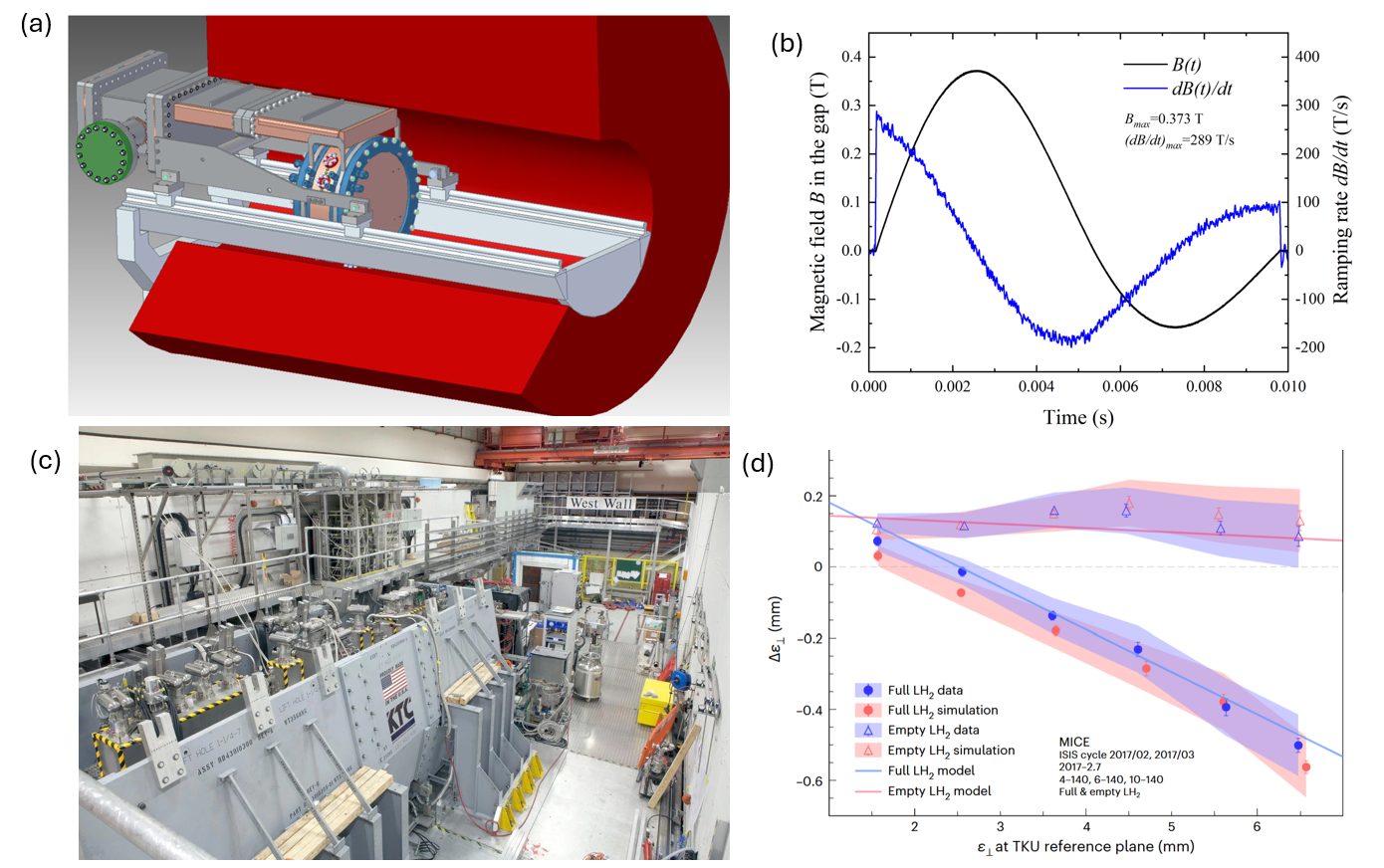}
\caption{Examples of MC accelerator R\&D: a) 50MV/m 805MHz RF cavity mounted in a 3 T  solenoid bore in the MTA facility at Fermilab~\cite{bowring2020operation, freemire2018gasfilled}. False colors indicate: cut view of MTA solenoid (red); support rails (silver); vacuum pumping port (dark green); RF pickup and instrumentation
ports (red); b) Fast 300T/s HTS magnet test results: magnetic field in the magnet gap $B(t)$ and the ramping rate $dB(t)/dt$ for a single pulse of the maximum current of about 1000 A in bipolar current pulse, from~\cite{piekarz2022record}; c) The Muon Ionization Cooling Experiment (MICE), pictured here at Rutherford Appleton Laboratory in the United Kingdom, has for the first time successfully cooled a beam of muons, essentially focusing a diffuse cloud of muon particles. The collaborators used 3 T superconducting focusing solenoids and liquid Hydrogen (LH$_2$) energy absorbers to demonstrate ionization cooling of muons; d) Emittance change between the upstream and downstream muon trackers of the MICE experiment, $\Delta \varepsilon_{\perp}$, as a function of upstream emittance for 140 MeV/c beams crossing 2.4 mm of the LH$_2$ MICE absorber. Results for the empty cases, "Empty LH$_2$" data, simulation, and model, are also shown. The measured effect is shown in blue, whereas the simulation is shown in red. The corresponding semitransparent bands represent the estimated total standard error. The error bars indicate the statistical error and for some of the points, they are smaller than the markers. The solid lines represent the approximate theoretical model  for the absorber (light blue) and empty (light pink) cases. The dashed gray horizontal lines indicate a scenario where no emittance change occurs (from~\cite{mice2024transverse}).}
\label{fig4}
\end{figure} 

Eventually, multiple Coulomb scattering limits further transverse emittance reduction at the initial operating momentum. The final cooling stage therefore operates at lower muon momenta $\sim 40\,\mathrm{MeV}/c$, where increased ionization energy loss enhances the cooling rate and stronger focusing reduces the transverse beta function. Dedicated high-field solenoids, reaching approximately $40\,\mathrm{T}$, provide the required focusing around the absorbers. This approach achieves additional transverse cooling at the expense of increased longitudinal emittance, which can be tolerated in the final stages. The resulting combination of low-energy operation and exceptionally strong solenoidal focusing is a distinctive feature of the MC cooling system.


The muon collider design study~\cite{accettura2025mcEPPSU} has addressed several key technological challenges. Cost- and power-efficient HTS solenoid designs have been developed, providing sufficiently large apertures to accommodate the RF cavities while withstanding the substantial mechanical stresses generated by the strong magnetic fields. An experimental program to validate and optimise these designs is now essential.


The 6D cooling cells require RF accelerating gradients of up to $\sim 30\,\mathrm{MV/m}$ in strong solenoidal magnetic fields of $B_z \sim 2$--$17\,\mathrm{T}$. A major challenge is RF breakdown, which can be exacerbated by the magnetic field. Without an external magnetic field, electrons emitted from a localized region of the cavity surface spread over a relatively large area. In a strong solenoidal field, however, they follow magnetic field lines and strike a concentrated region of the opposite surface, potentially causing local damage and triggering further breakdowns. This challenge is being addressed through improved cavity designs, including the use of beryllium windows. Previous experiments demonstrated accelerating gradients exceeding $50\,\mathrm{MV/m}$ in a magnetic field, as shown in \textbf{Figure~\ref{fig4}a}. Since the original test facility has been dismantled, establishing new high-gradient RF test stands is essential for further development and validation.

A second challenge arises in the final cooling stages, where the highly concentrated muon beam can damage solid absorbers. Hydrogen-vapor-filled absorber designs offer a promising alternative by distributing the beam energy deposition throughout the gas. Initial tests of the containment windows have successfully demonstrated their ability to withstand proton beams producing greater mechanical stresses than those expected from the muon beam. Further development and testing will be needed to optimise the robustness and reliability of these designs.



\subsection{Muon Acceleration Systems}

\begin{marginnote}[]
\entry{RCS}{Rapid Cycling Synchrotron, a specialized circular particle accelerator designed to ramp its magnetic fields up and down at extremely high speeds}
\end{marginnote} 

The muons are accelerated in a sequence of a linac, two recirculating linacs and several pulsed synchrotrons (RCS, which stands for rapid cycling synchrotron). The focus of the studies was on the chain of pulsed synchrotrons, the most important cost item and power consumer of the complex.

The synchrotrons must ramp rapidly to minimize muon decay losses. The acceleration time ranges from approximately $0.4\,\mathrm{ms}$ in the first RCS to $10\,\mathrm{ms}$ in the last. Such rapid cycling currently requires normal-conducting magnets, as conventional superconducting magnets cannot sustain the required ramp rates. However, field ramping rates of $\sim 500-1000\,\mathrm{T/s}$, sufficient for the final and slowest RCS, may become achievable with HTS dipole magnets~\cite{piekarz2022record}, $\sim 300\,\mathrm{T/s}$ has already been demonstrated for smaller amplitudes as illustrated in \textbf{Figure~\ref{fig4}b}.

Normal-conducting dipole magnets are limited to fields of approximately $1.8\,\mathrm{T}$ while maintaining adequate field quality. Consequently, the final RCS, which accelerates muons to $5\,\mathrm{TeV}$ before injection into the superconducting collider ring, would require a very large circumference. Its size can be substantially reduced by interleaving static superconducting dipoles with fast-ramping magnets, with the static magnets providing the larger share of the integrated bending field~\cite{rcs025optics}. At injection, the ramping magnets operate at a negative field, $-B$, partially compensating the static bending field. As the beam energy increases, their field is ramped from $-B$ to $+B$, progressively adding to the static field. This arrangement approximately halves both the required length of fast-ramping magnets and the circumference of the final RCS. Nevertheless, at about $35\,\mathrm{km}$, it remains the largest ring in the MC complex, compared with approximately $11.4\,\mathrm{km}$ for the 10 TeV collider ring.


The pulsed synchrotrons pose beam dynamics and technical challenges~\cite{accettura2025mcEPPSU}. Lattice designs have been developed for all RCSs. Studies of the impedance effects and the beam-based feedback show that the very intense bunches will remain stable.

The total energy in the field of the fast-ramping magnets reaches about $200\,\rm MJ$. If the energy was lost at each pulse more than 1 GW would be required to drive the facility. Power converter can recover the energy from one magnet ramp into capacitors to be reused for the next pulse. In a simple design, the magnets that invert their field would also invert the voltage in the capacitors, dramatically reducing their lifetime. A power converter design has been developed that avoids this voltage inversion, leading to an important reduction of the system cost.

Another critical challenge is to synchronize the increase in beam energy with the magnetic field ramp, maintaining the required proportionality $E_\mu(t) \propto B(t)$ to ensure a stable beam orbit and preserve beam quality. The RF system generally favors a linear increase in beam energy with time, whereas the magnets and their power converters favor approximately sinusoidal field variations, as in a resonant circuit. Enforcing a linear field ramp would require prohibitively expensive pulse-forming networks. A system-level optimization accounting for the interplay among beam dynamics, RF acceleration, magnets, and power converters has therefore led to a cost- and power-efficient design. Since losses in the magnets and power converters are a major contributor to the overall power consumption, the optimized design limits magnet-system energy losses to approximately $1$--$4\%$ per pulse, depending on the RCS.


Studies of the RF systems considered the strong beam loading from the high bunch charge and developed the feedback required for stable operation. The studies show that $1.3\,\rm GHz$ superconducting cavities can be used and that the beam quality can preserved during the acceleration in the chain of the RCSs.

\subsection{Collider Ring Design}

The collider ring stores the two counter-rotating muon bunches and brings them into collision at two IPs on opposite sides of the ring. At 10 TeV, high-field HTS dipole magnets are envisaged to minimize the ring circumference. For the initial 3 TeV stage, however, this technology may not yet be sufficiently mature for accelerator dipoles, which present greater technical challenges than solenoids. The baseline design therefore assumes $\mathrm{Nb_3Sn}$ dipoles, based on technology developed for the HL-LHC. Strong quadrupole magnets on either side of each IP focus the beams to extremely small transverse sizes at the collision point, maximizing luminosity.


A major challenge for the collider ring arises from muon decays. Each muon decays into an electron or positron and two neutrinos, and  
the energy spectrum
of the electrons and positrons is very wide with an average of about one third of the muon energy.
Because of their lower magnetic rigidity, these particles follow more strongly curved trajectories and strike the inner wall of the beam pipe. They also emit energetic synchrotron radiation photons, which strike the outer wall. The resulting power deposition is approximately $500\,\mathrm{W/m}$, an unacceptable heat load for superconducting magnets. Removing $1\,\mathrm{W}$ of heat at $4.5\,\mathrm{K}$ requires approximately $300\,\mathrm{W}$ of electrical power, implying a total cryogenic power consumption of about $1.5\,\mathrm{GW}$ without shielding. Operation of HTS magnets at $20\,\mathrm{K}$ would reduce this demand, but it would remain substantial. Detailed FLUKA simulations have guided the design of a $3\,\mathrm{cm}$ thick tungsten-based shield, which reduces the heat load on the magnets by approximately two orders of magnitude. Operating near room temperature, the shield can be cooled efficiently, keeping the total cooling power below $10\,\mathrm{MW}$. The shielding also protects the superconducting magnets from radiation damage.


The collider dipole magnets must combine the highest possible magnetic fields with large apertures to accommodate the shielding. The resulting increased electromagnetic forces make their design particularly challenging. Optimized HTS dipole designs, balancing these competing requirements, achieve fields of $14\,\mathrm{T}$ with an aperture of $14\,\mathrm{cm}$. However, HTS dipole technology is less mature than that of HTS solenoids. For the initial 3 TeV collider stage, the baseline design therefore relies on $\mathrm{Nb_3Sn}$ technology developed for the HL-LHC upgrade. This approach requires thicker shielding and limits the achievable dipole field to approximately $11\,\mathrm{T}$.


The collider ring optics presents significant challenges~\cite{alexahin2018muon}. The beams must maintain their short bunch lengths while being focused to extremely small transverse sizes at the IPs. Such strong focusing causes the beam size to increase rapidly away from the IPs, requiring large-aperture final-focus quadrupoles and complicating their design. It also introduces substantial chromatic aberrations, analogous to those in optical lenses, which limit the momentum acceptance of the collider ring and thereby affect the achievable luminosity (see {\textbf {Equation~\ref{lumi}}).

The initial optics design~\cite{accettura2025mcEPPSU} approaches the target performance, although further improvements in momentum acceptance are needed to fully meet the design goals. Studies of collective effects, particularly beam coupling impedance and beam--beam interactions, indicate that they do not significantly compromise the expected performance.

\begin{figure}[h]
\includegraphics[width=5in]{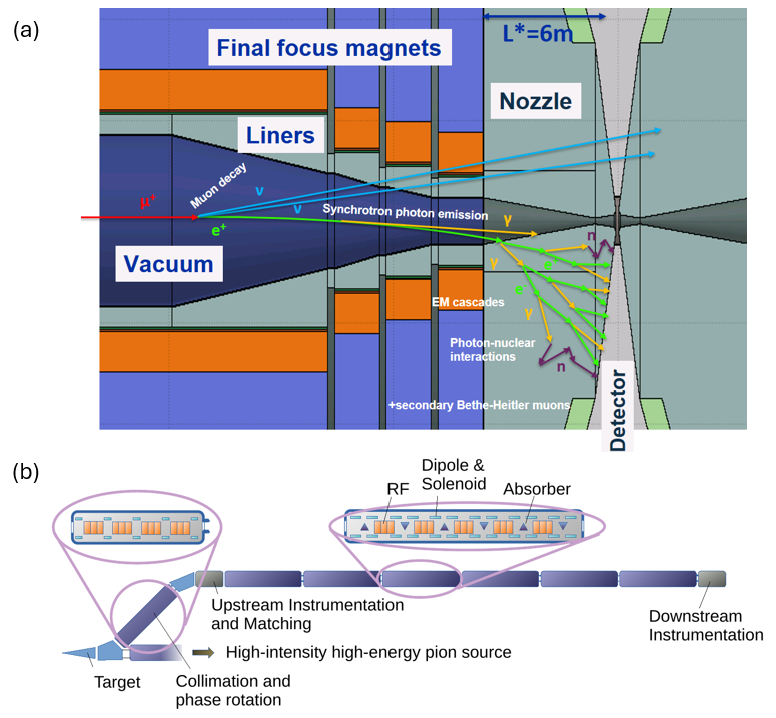}

\caption{
(a) Schematic illustration of beam-induced backgrounds from muon decays in the IR and adjacent magnets. A muon (red) decays into a positron (green) and two neutrinos (blue). Deflected by the magnetic fields, the positron radiates high-energy photons (yellow) and loses energy. Both the positron and the emitted photons initiate electromagnetic showers upon striking the beam vacuum chamber walls, producing $e^\pm$, $\gamma$, secondary $\mu^\pm$, and neutrons (violet). Most of these secondary particles are intercepted by the tungsten shielding nozzles inside the detector;
(b) Conceptual layout of the muon cooling demonstrator: multi-GeV protons hit a target and produce muons; dipoles, RF cavities, and collimators select and prepare a muon beam with momenta in the range 190--210~MeV/$c$. After beam characterization and matching, the muons pass through an ionization cooling channel comprising solenoids, dipoles, RF cavities, and absorbers, followed by measurements of the cooled beam. From Ref.~\cite{rogers2023demonstrator}.}
\label{fig5}
\end{figure}

\subsection{Radiation Protection}

Environmental protection is a key design consideration for the muon collider study including radiation protection.
Conventional radiation protection issues, including prompt and residual radiation, activation of surrounding materials, and radioactive waste, are shared with other collider facilities and can be addressed through appropriate shielding and facility design. 
The self-imposed goal is to limit the total level to less than $10\,\mu\mathrm{Sv}$ per year, the level that CERN achieves. This is much below the legal limit and less than 1\% of the ICRP reference values of $1\,\rm mSv$ per year for human-caused radiation (which excludes medical exposure). Studies show that the muon collider can achieve this goal.

To achieve this goal a distinctive feature of the muon collider needed to be addressed, a dense very narrow neutrino flux produced by muon decays in the collider ring. Neutrinos interact weakly with matter, but a very small fraction can interact in the ground near the Earth's surface and generate secondary localized particle showers that exit the surface. The effect is confined to small predictable surface areas many kilometers away from the collider.

Detailed FLUKA simulations have been used to develop a neutrino radiation dose model taking into account beam energy, collider lattice, neutrino angular distributions, and realistic exposure scenarios. The highest localized doses appear in the spots that the two straights housing the detectors point towards; the highest dose rate corresponds to that in an airplane. Careful orientation of the collider ring points these straight to uninhabited land that can be fenced on the one side and to the Mediterranian sea far away on the other, which further reduces the dose.
A mechanical system slowly moves the magnets of the arcs and changes the beam orbit widening the very narrow cone of $20\,\rm\mu radian$. Simulations indicate that this approach can reduce peak neutrino-induced doses by factors of $\sim100$~\cite{carli2023jacow}.

With these measures the goal can be met. Activation of the soil and ground water are negligible and the dose to plants and animals are far below the ICRP reference values.

\subsection{Synergies with Other Facilities}

Muon collider technologies offer significant synergies with other fields. Large-aperture, 20~T HTS target solenoids are relevant to fusion energy, while muon cooling solenoids, ranging from large-aperture 6D cooling magnets to small-aperture, 40~T final-cooling magnets, have potential applications in medical technologies, materials science, and high-field research. High-power targets are also essential for spallation neutron sources, such as SNS and ESS, and neutrino facilities. In addition, muon decays in the injector complex and collider ring could provide neutrino beams for dedicated experiments.


\section{EXPERIMENTAL ENVIRONMENTS AND DETECTORS}
\label{Detector}

The experimental environment of a high-energy MC is dominated by beam-induced backgrounds (BIB) from muon decays, imposing demanding requirements on the machine--detector interface (MDI), shielding, detector technologies, and event reconstruction. Effective background rejection exploits differences in particle energies, arrival times, and directions. Tracking algorithms use timing and trajectory consistency with the interaction point, while calorimetry relies on timing, cluster shapes, and spatial correlations. Fine detector granularity enables particle-flow reconstruction, complemented by advanced pattern-recognition and machine-learning techniques.

A comprehensive discussion, including detailed simulations and extensive references, can be found in Ref.~\cite{Casarsa2024}. Here, we focus on the principal challenges, representative detector parameters, and approaches being developed to address them.

\begin{marginnote}[]
\entry{BIB}{Beam-Induced Background is the stream of unwanted secondary and tertiary particles entering a detector, caused by beam particles colliding with accelerator components, residual gas, or decaying while in transit}
\entry{MDI}{Machine Detector Interface is the boundary region where the accelerator meets the detector, integrating final focus magnets, shielding, and the beam pipe around the interaction point to maximize collision luminosity while protecting the experiment from BIB.}
\entry{IR}{Interaction Region is the specific section of a collider where beams collide, encompassing the IP, final focus magnets, and surrounding detector subsystems.}
\end{marginnote}

\subsection{Beam-Induced Backgrounds}
\label{BIB1}

The dominant source of BIB at a MC is the decay of stored muons. At 10 TeV, approximately 58,000 muons decay per meter of beamline per bunch passage, producing energetic electrons and positrons. While decay products generated near the IP largely follow the beams out of the detector, those produced farther upstream can be deflected by the strong interaction-region magnets or strike accelerator components, generating secondary electromagnetic and hadronic showers. The resulting BIB includes photons, electrons, neutrons, charged hadrons, and secondary muons, with broad energy and arrival-time distributions.

An additional background arises from incoherent pair production (IPP), $\mu^+\mu^- \rightarrow \mu^+\mu^- e^+e^-$, during collisions at the IP. Although smaller in overall particle flux, IPP is an important source of high-momentum background electrons and positrons entering the detector.

Early US MAP studies established the importance of jointly optimizing the interaction-region optics, superconducting magnets, and machine--detector interface (MDI)~\cite{Mokhov2011IR,Mokhov2012Background}. The principal protection against decay-induced backgrounds is provided by conical shielding nozzles surrounding the incoming and outgoing beams, supplemented by tungsten masks. The nozzles, extending approximately 6 m from the IP with an opening angle of about $10^\circ$, consist primarily of tungsten with borated polyethylene for neutron moderation and capture. Their geometry is optimized to suppress BIB while preserving forward detector acceptance~\cite{accettura2023towards,Casarsa2024}. \textbf{Figure~\ref{fig5}b} illustrates the background-generation and shielding principles.

Detailed particle-transport simulations are essential for evaluating detector occupancies and radiation loads. At a 10 TeV MC, the estimated ionizing radiation doses and neutron fluences are broadly comparable to those at the HL-LHC, implying similar radiation-hardness requirements despite differences in background composition, timing, and occupancies~\cite{Casarsa2024,MuonColliderForum2024}.

The much lower bunch-crossing rate of a MC, approximately 100 kHz compared with 40 MHz at the HL-LHC, provides more time for event readout and processing. Effective BIB suppression relies on precise hit timing (typically below 100 ps), directional discrimination using double-layer vertex detectors, pulse-shape analysis, and hit-cluster characteristics. Recent advances in reconstruction techniques and tracker-layer optimization have demonstrated that some of these requirements, particularly the stringent timing resolution, can be relaxed without compromising background rejection.

\subsection{Detector Concepts and Key Requirements}

Two multipurpose detector concepts, MUSIC (MUon System for Interesting Collisions) and MAIA(Muon Accelerator Instrumented Apparatus) have recently been developed for a muon collider operating at a center-of-mass energy of 10 TeV~\cite{musics2026detector, maia2025detectors}. Both build on earlier detector studies while incorporating substantial modifications to accommodate the higher collision energy and the challenging BIB. They feature large-volume silicon tracking systems, highly segmented electromagnetic and hadronic calorimeters, 5 T solenoids, and external muon identification systems. Their designs aim to maintain high reconstruction efficiency and momentum resolution while suppressing backgrounds from muon decays and incoherent electron--positron pair production.

The tracking systems of both detectors comprise silicon pixel vertex detectors surrounded by inner and outer trackers. The vertex detectors employ $25\times25\,\mu\mathrm{m}^2$ pixels, with assumed hit timing resolutions of approximately 30 ps, while the larger-radius tracking layers use silicon macropixels with timing resolutions of about 60 ps. MUSIC features five barrel vertex layers, whereas MAIA uses four, retaining double-layer structures where needed for background rejection and track reconstruction. Both concepts employ superconducting solenoids providing a central magnetic field of $5\,\mathrm{T}$, with inner diameters and lengths of approximately $4\times5\,\mathrm{m}^2$ for MUSIC and $3\times4.6\,\mathrm{m}^2$ for MAIA. The strong magnetic field improves the momentum resolution of high-energy charged particles and reduces the impact of low-momentum backgrounds.

The principal differences between MUSIC and MAIA lie in their calorimeter technologies and placement. MUSIC employs a lead-fluoride-crystal-based electromagnetic calorimeter inside the solenoid, while MAIA uses a silicon--tungsten sampling calorimeter outside the solenoid. Both concepts feature iron--scintillator hadronic calorimeters and external muon identification systems.

Despite substantial BIB, initial studies indicate that the relatively long intervals between bunch crossings could enable continuous, streaming data acquisition. The detectors are therefore designed for continuous operation without power pulsing, using circulating gas for cooling.

Detailed simulations of both detector concepts, incorporating backgrounds from muon decays and incoherent pair production, have demonstrated promising performance in tracking, calorimetry, and reconstruction of charged particles, photons, leptons, hadronic jets, and Higgs bosons, including flavor tagging~\cite{Casarsa2024,MuonColliderForum2024}. Further R\&D will focus on fast, radiation-tolerant silicon sensors, calorimeter technologies, background rejection, and reconstruction algorithms, as well as demonstrating that all detector performance requirements can be met simultaneously. Engineering development must also address mechanical supports, cooling, power distribution, and subsystem integration.

\subsection{Forward Physics and Additional Opportunities}

The forward region presents both important physics
opportunities and substantial engineering challenges.
The shielding nozzles and nearby final-focus magnets
restrict acceptance at small polar angles, affecting
measurements involving forward muons, missing
momentum, and vector-boson fusion processes.
Forward-muon tagging is particularly valuable
for VBF measurements, including studies
of the Higgs boson and its total decay width.
Improved acceptance could substantially enhance
these measurements, but extending instrumentation
toward the beamline must be balanced against
background exposure, shielding requirements,
and machine protection
\cite{Casarsa2024,accettura2025mcEPPSU}.
Additional opportunities may arise from the intense
neutrino fluxes generated by muon decays, potentially
supporting dedicated neutrino and nuclear physics experiments~\cite{neutrino2025muoncoll, burk2026neutrino}. Such
measurements would require specialized detector
locations and shielding arrangements rather than
extensions of the central collider detector. \\

In summary, studies initiated within the US MAP and continued by the IMCC and USMCC have demonstrated that high-quality physics measurements are achievable in the challenging MC environment using near- to medium-term detector technologies. This progress builds on advances in radiation-hard silicon sensors, precision timing, highly segmented calorimetry, and modern reconstruction algorithms.  The advanced multipurpose detector concepts MUSIC~\cite{musics2026detector} and MAIA~\cite{maia2025detectors} have been developed for a 10 TeV muon collider and are also adaptable to the initial 3 TeV stage. The next priorities are to demonstrate that all detector performance requirements can be met simultaneously and to develop practical engineering solutions for mechanical supports, cooling, power distribution, and subsystem integration.


\section{Outlook}
\label{Outlook}
 
The international high-energy physics community stresses the importance of reaching
parton-parton centre-of-mass energies around $10\,\rm TeV$. This could be achieved with
a $\sim$100$\,\rm TeV$ hadron collider or a 10$\,\rm TeV$ lepton collider.
Recent strategy planning exercises in Europe and the US compared options for future colliders
in terms of physics, cost, power consumption, technological readiness, timeline and
environmental impact\cite{roser2023itf, Arduini2025ColliderComparison}.
In Europe a 10~TeV muon collider, with a potential initial stage at 3~TeV, is considered a
long-term option to follow an FCC-ee Higgs factory,
the other being the FCC-hh~\cite{ESG2026Strategy}.
In the US, the community called for the exploration of a 10 TeV muon collider facility and the 2023 Particle Physics Project Prioritization Panel (P5) recommended to explore
the possibility to host it, calling it the  ``muon shot'' for the future of US particle
physics,~\cite{P5Report2023}. In 2025, the
National Academies Report emphasized the importance of hosting such a facility by the
mid of the century and made the support for the R\&D its highest
priority~\cite{NASEM2025ParticlePhysics}.

\subsection{Collider Implementation}
IMCC focuses on a greenfield configuration that allows exploration
of the physics potential as well as the challenges and limitations of the
design and the technologies. Exploratory studies of specific sites have started
at CERN and Fermilab.

For the European Strategy, studies of the CERN layout have been performed that focus on an option with the whole surface buildings of the complex on CERN-owned land and minimising civil construction.
The existing SPS would house the first RCS and the LHC tunnel one
RCS for a 3.2 TeV option and two RCSs for 7.6 TeV. Only an additional collider ring
needs to be constructed, with less than 10 km circumference for the
highest energy. \textbf{Figure~\ref{fig6}d} shows the layout and tunnel cross section.
At Fermilab, the feasibility has been explored to accommodate the facility within the
existing laboratory site and reuse parts of the proton accelerator complex
for muon production.

Studies of the environmental impact of a MC at CERN indicate that radiation levels can be kept negligible, comparable to those currently achieved at the LHC complex.
This relies on the arc mover system and on pointing the experimental insertions to a site in the Jura, which can be used for neutrino experiments. Similar studies at Fermilab have started.

Ultimately, the IMCC, including the USMCC and with the broader international high-energy physics community, aim to develop a global consensus on the preferred collider configuration, host site, and implementation strategy.

\begingroup
\renewcommand{\arraystretch}{1.5} 
\begin{table}[h]
\tabcolsep7.5pt
\caption{Integrated luminosity per year summed over all experiments (two in MC, one in CLIC, four in FCC-hh) assuming $\sim 10^{7}$\,s of operation per year, time to first physics (technical limit from the decision to proceed), construction cost (2021 B\$, parametric estimate without escalation and contingency),
peak electric power for green-field MCs and competing collider projects. Adapted from~\cite{roser2023itf}.}
\label{TableITF}
\begin{center}
\begin{tabular}{@{}lc|c|c|c|c@{}}
\hline
Collider & & ${\cal L}_{\rm int}$/yr. & Time to & Cost Range & Electric power \\
 & & ab$^{-1}$/yr & 1st physics & 2021 B\$ & MW \\
\hline
3 TeV MC & $\mu^{+}\mu^{-}$ & 0.4 & 19-24 & 7--12 & 117 $^{\rm a}$ \\
10 TeV MC & $\mu^{+}\mu^{-}$ & 4 & $>$25 & 12--18 & 182 $^{\rm a}$ \\
\hline
3 TeV CLIC & $e^{+}e^{-}$& 0.6 & 19-24 & 18-30 & $\sim$550 \\
100 TeV FCC-hh & $pp$ & 6 & $>$25 & 30-50 & $\sim$560 $^{\rm b}$ \\
\hline
\end{tabular}
\end{center}
\begin{tabnote}
$^{\rm a}$ Based on detailed estimates in
Ref.~\cite{accettura2025mcEPPSU};
$^{\rm b}$ 
Following the FCC Feasibility Study~\cite{benedikt2025fcc}, the FCC-hh parameters have been revised relative to those in the ITF report. The current proposal envisions an 85 TeV proton-proton collider following FCC-ee, with first physics in 2074, a peak power consumption of 355 MW, and annual electricity consumption of 2.45 TWh.
\end{tabnote}
\end{table}
\endgroup

\subsubsection{Timeline, Staging, and Prospects}
The realization of a high-energy MC will require
a sustained international R\&D effort, followed by the design,
construction, and commissioning of a complex accelerator facility.

For the 2025 European Strategy process the collaboration has developed a technically limited,
success-oriented roadmap that could enable first
collisions by the mid-century, provided sufficient resources
are available and the principal technical challenges
are resolved~\cite{accettura2025mcEPPSU}. The collaboration has developed a similar timeline for an implementation in the US
that considers some limitations for funding\cite{usmcc2025eppsu}.
The roadmaps begin with a technology R\&D phase of approximately seven years, aimed at establishing the feasibility of key accelerator systems. This is followed by a seven to ten year technology demonstration phase, including the design, construction, and operation of a 6D muon cooling demonstrator---see {\textbf Figure \ref{fig5}b}--- and leads to technical design reports with detailed cost estimates to enable a potential decision on the construction.
Following successful demonstration of muon cooling, a five-year project preparation phase would complete the engineering design, technology development and industrialization, as well as site implementation studies and authorization procedures. The subsequent MC construction phase, encompassing civil engineering, infrastructure installation, accelerator construction, and commissioning, would take approximately nine years.

The key R\&D priorities are discussed in the
following subsection.  Under this scenario, a MC operating at
approximately 3 TeV center-of-mass energy could
begin operation several years after
the conclusion of the HL-LHC program and it is an attractive candidate to
complement a Higgs factory. This ambitious timeline represents a technically feasible
scenario rather than a detailed construction plan or a commitment
to specific dates.

\subsubsection{Cost and Power Consumption Estimates}

The Snowmass 2021 international Collider Implementation Task Force (ITF) compared future collider proposals and concepts in terms of luminosity, environmental impact, power consumption, R\&D duration, time to first physics, and cost. Its report~\cite{roser2023itf} identifies a multi-TeV MuC as a particularly promising option for reaching the 10+~TeV energy frontier, with one of the lowest estimated cost ranges among the concepts considered (see Table~\ref{TableITF}).

A detailed cost estimate has been made for the CERN and a greenfield site. It is based on the conceptual
designs of the key components and on detailed scaling from other projects for more
generic components~\cite{accettura2025mcEPPSU}.
The cost ranges are shown in~\textbf{Figure~\ref{fig6}a} and
the distribution of the cost for the 7.6~TeV case, in \textbf{Figure~\ref{fig6}b}.
These bottom-up estimates are very consistent with the prior ITF estimates.
An assessment of the cost uncertainties is shown in \textbf{Figure~\ref{fig6}c}.
A systematic review assessed the potential
to reduce the cost and its uncertainty with further advances in the technologies.


The electrical power demand of a 10~TeV muon collider is estimated at approximately 200~MW, with preliminary power budgets allocating about 10\% to the proton driver, 6\% to muon cooling, 5\% to recirculating linacs, 65\% to rapid-cycling synchrotrons, 3\% to the collider ring, and 11\% to supporting infrastructure~\cite{accettura2025mcEPPSU}.

Accounting for operational cycles, standby periods, and downtime, the projected annual electricity consumption is approximately 1.2 TWh, comparable to CERN's current annual consumption, of which about 90\% is attributable to its accelerator complex.

These estimates are substantially lower than those
for other proposed energy-frontier colliders, see \textbf{Table~\ref{TableITF}}, highlighting the potential
of muon colliders to provide an energy-efficient
path toward the 10 TeV parton center-of-mass
energy frontier and even beyond.

\begin{figure}[h]
\includegraphics[width=5in]{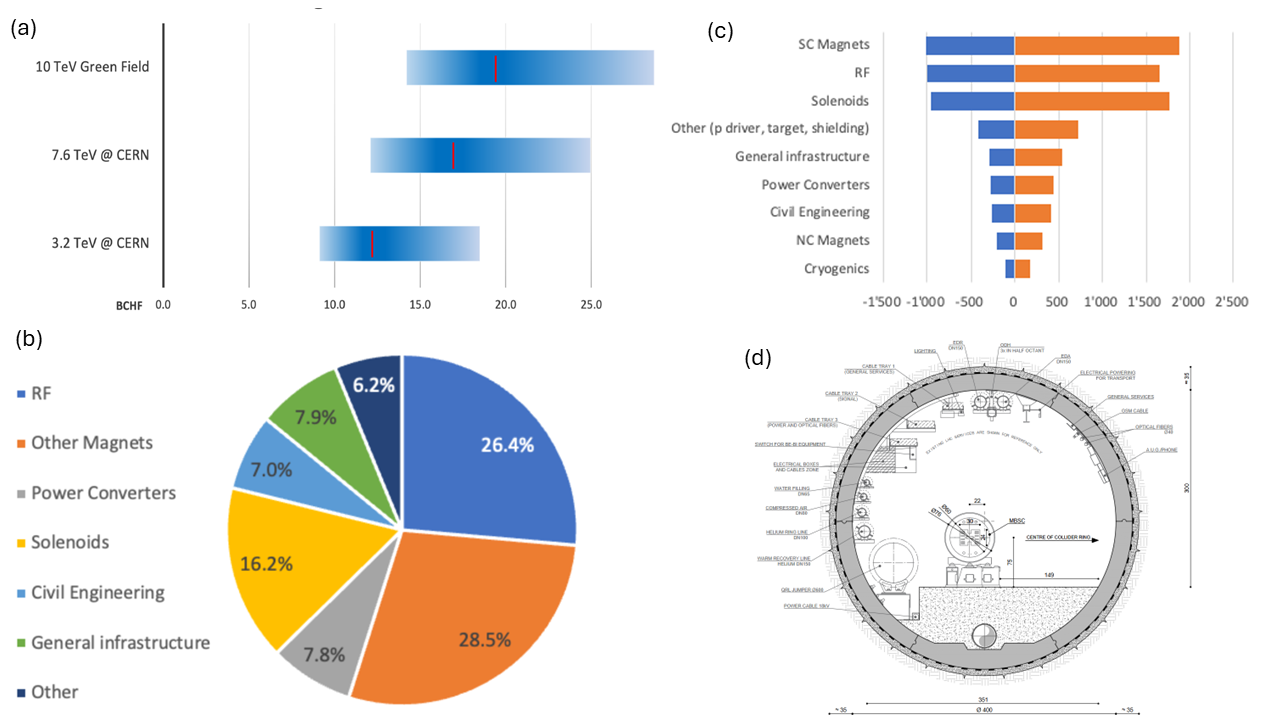}

\caption{Cost estimates and infrastructure layout for a muon collider:
(a) Cost distributions for CERN site-specific MC configurations and a 10~TeV greenfield facility (in BCHF). The darkest portions of the color bands indicate the most probable costs, while the lighter portions represent progressively less probable values, reflecting the extent of the calculated uncertainty; red markers identify the most probable costs (the cost of the two experiments is not included in the estimate);
(b) Percentage distribution of costs by technology for a CERN-based 7.6~TeV muon collider;
(c) cost uncertainties (in MCHF) by technology for the same facility, with blue bars indicating potential cost reductions (negative variance) and orange bars indicating potential cost increases (positive variance);
(d) cross-sectional arrangement of components in a muon collider tunnel with a 4~m inner diameter.
From Ref.~\cite{accettura2025mcEPPSU}.}
\label{fig6}
\end{figure}

\subsection{Technical Demonstrators and R\&D Roadmap}
Building on recent progress, the IMCC has developed a coordinated R\&D program with defined deliverables, resource estimates, and milestones toward a Conceptual Design Report (CDR). The program has four principal components:
\begin{itemize}
\item Development of an integrated end-to-end accelerator design, optimized for cost, power consumption, and performance, supported by comprehensive simulation tools and cost and power models.
\item A staged experimental program to develop and demonstrate ionization cooling.
\item Experimental validation of high-temperature superconducting (HTS) magnet and RF technologies.
\item Further development of detector technologies and reconstruction methods to minimize the impact of beam-induced backgrounds.
\end{itemize}

\begin{marginnote}[]
\entry{CDR}{Conceptual Design Report, outlining the proposed facility, its feasibility, and preliminary design}
\entry{TDR}{Technical Design Report, detailing the engineering design and specifications for construction}
\entry{TRL}{Technology Readiness Level, a scale from 1 to 9 assessing technology maturity, from basic principles (TRL 1) to proven operation (TRL 9); TRL 6 denotes prototype demonstration in a relevant environment}
\end{marginnote}

The program comprises 51 deliverables, with estimated resources of approximately 320~MCHF in materials and 2700~FTE-years for the collider and detector R\&D part over a ten-year time period~\cite{accettura2025mcEPPSU}. Muon cooling demonstrations and magnet development dominate both the resources and schedule. Before proceeding to the Technical Design Report (TDR) phase, the magnets are expected to reach Technology Readiness Level (TRL)~6, and the cooling demonstrator must successfully complete beam tests. 

An initial design of a muon cooling demonstrator~\cite{rogers2023demonstrator}, shown in \textbf {Figure \ref{fig5}b}, has been developed for potential implementation at CERN, while similar studies are underway in the US~\cite{jindariani2026scac}. The demonstrator facility is considered potentially synergistic with an entry-level neutrino factory~\cite{neutrinofact2025modern}, a concept historically envisaged as sharing the front-end of a MC~\cite{geer2009muon}. 

As initial steps toward a cooling module prototype, SLAC is constructing an RF test stand to evaluate cavity performance in strong magnetic fields, while INFN is planning a split HTS solenoid to test representative cooling-channel field configurations.

Given current resource constraints, the European roadmap for 2027--2031 spreads the initial three to four years of the full R\&D program over a longer period while preserving the possibility of faster implementation. Apart from the HTS split-coil system, magnet activities will initially focus on design studies and collaborative research with partners.

\section{Summary}

Over the past decade and a half, studies of energy-frontier muon colliders by the US MAP, IMCC, and USMCC have achieved substantial progress, with no fundamental showstoppers identified. Advances in muon production and cooling, rapid acceleration, high-field magnets, collider ring design, and machine--detector integration have established a credible technical foundation for 3 and 10 TeV colliders. Preliminary cost and power estimates indicate promising prospects for a compact and energy-efficient facility, while initial site studies have yielded encouraging results. The next decisive step is a coordinated international R\&D program, including hardware development and experimental demonstrations of critical technologies, particularly six-dimensional muon cooling. Sustained efforts over the next 10--15 years could lead to a Technical Design Report and pave the way for construction of an energy-frontier muon collider by mid-century.

Innovation and cutting-edge R\&D are central to this endeavor. The technologies required for a muon collider offer important synergies with other areas of particle physics and broader applications, ranging from medical technologies to fusion energy. Equally important, the project offers exceptional opportunities for early-career researchers and engineers to pursue ambitious ideas and help shape the future of particle physics.
The participation of more than 300 early-career researchers in recent IMCC and USMCC annual meetings, together with their active involvement in muon collider R\&D and high-energy physics strategic planning in the US and Europe, demonstrates the strong appeal of the collider's scientific goals and technological challenges. A sustained international effort will not only bring the muon collider closer to realization but also inspire and prepare the next generation of scientists and engineers who will build it.

\section*{SIGNIFICANCE STATEMENT}

Particle colliders are among humanity's most powerful tools for exploring the fundamental laws of nature. They have revealed the basic building blocks of matter, confirmed the Standard Model of particle physics, and led to the discovery of the Higgs boson. However, reaching higher energies with existing collider designs is becoming increasingly difficult because future machines would require enormous size, cost, and power.

A muon collider offers a promising new approach, combining very high collision energies with the potential for a more compact, cost-effective, and energy-efficient facility. Recent advances in accelerator science and technology, together with focused design studies of 3--10~TeV muon colliders, have brought this concept closer to reality. This review explains the scientific opportunities, technological challenges, and research needed to pave the way for a muon collider to become the next great discovery machine by mid-century.

\section*{DISCLOSURE STATEMENT}
The authors are not aware of any affiliations, memberships, funding, or financial holdings that
might be perceived as affecting the objectivity of this review. 

\section*{ACKNOWLEDGMENTS}
We would like to thank the many people who contributed with their hard work, useful discussions, and valuable advice: Sergo Jindariani, Patrick Meade, Diktys Stratakis, Chris Rogers, Lucio Rossi and others. We are grateful to the International Muon Collider Collaboration, the US Muon Accelerator Program and the US Muon Collider Collaboration
for their support. We acknowledge the financial support of  the European Organization for Nuclear Research (CERN). This work was supported via the MuCol Design Study by the European Union (EU). Views and opinions expressed are however those of the authors only and do not necessarily reflect those of the EU or European Research Executive Agency (REA). Neither the EU nor REA can be held responsible for them.

%

%
%
%
%
%
%
%
%


\bibliographystyle{apsrev}
\bibliography{ARMCrefs}

%
%
%
%
%
%
%
%
%
%
%
%
 
\end{document}